# An Investigation on Non-Invasive Brain-Computer Interfaces: Emotiv Epoc+ Neuroheadset and Its Effectiveness

Md Jobair Hossain Faruk
*Dept. of Software Engineering and Game Dev.*
*Kennesaw State University*
Marietta, USA
mhossa21@students.kennesaw.edu

Maria Valero
*Dept. of Information Technology*
*Kennesaw State University*
Marietta, USA
mvalero2@kennesaw.edu

Hossain Shahriar
*Dept. of Information Technology*
*Kennesaw State University*
Marietta, USA
hshahria@kennesaw.edu

*Abstract*—Neurotechnology describes as one of the focal points of today's research of hundreds of academicians and researchers globally; particularly around the domain of Brain-Computer Interfaces (BCI) that was introduced in the 1970s and has been at the forefront of many neurotechnological discoveries. The primary attempts of BCI research are to decoding human speech from brain signals, implementing creativity by imagination, and controlling neuro-psychological patterns by utilizing billion of neural activities that would significantly benefit people suffering from neurological disorders. In this study, we illustrate the progress of BCI research and present scores of unveiled contemporary approaches. First, we explore a decoding natural speech approach that is designed to decode human speech directly from the human brain onto a digital screen introduced by Facebook Reality Lab and University of California San Francisco. Then, we study a recently presented visionary project to control the human brain using Brain-Machine Interfaces (BMI) approach. We also investigate well-known electroencephalography (EEG) based Emotiv Epoc+ Neuroheadset to identify six emotional parameters including engagement, excitement, focus, stress, relaxation, and interest using brain signals by experimenting the neuroheadset among three human subjects where we utilize two two supervised learning classifiers, Naïve Bayes and Linear Regression to show the accuracy and competency of the Epoc+ device and its associated applications in neurotechnological research. We present experimental studies and the demonstration indicates 69 % and 62 % improved accuracy for the aforementioned classifiers respectively in reading the performance matrices of the participants. We envision that non-invasive, insertable, and low-cost BCI approaches shall be the focal point for not only an alternative for patients with physical paralysis but also understanding the brain that would pave us to access and control the memories and brain somewhere very near.

## I. INTRODUCTION

The past decade has seen a tremendous explosion of research in neurotechnology by adopting state-of-the-art technology, AI, and machine learning that helps to improve the ability of researchers to conceptualize the Brain-Computer Interfaces (BCI) [1]–[4]. One of the ambitions of the scientists is to pour the way of technology in conjunction with the brain that would lead them translating the speech by utilizing brain activities in real-time application [5]. The human brain contains about 100 billion neurons or nerve cells and responsible to receive sensory input from outer environments for transmitting information to other organs of the human body to control from simple survival to complex cognitive functions [6]–[8]. Such functions make the brain a foremost entity of the human body for every thought, action, memory, feeling, and everything related to innermost activities.

Despite the wealth of modern technology, no individual or mechanism can control people's desires, actions, feelings, thoughts, and values [9]. However, the goal of futuristic technologies is to control one's brain activities using not only smart devices, but also emotional parameters of the human being [10]–[12]. The idea of BCI was introduced by Professor Vidal [13] in 1973 with a project that presented the first systematic attempt to conceptualize the baselines of direct brain-computer communication using electroencephalography (EEG). EEG research can be described as one of the oldest neuro-scientific techniques, with the first human brain recordings published by Hans Berger in 1929 [14]. The EEG approach has matured over the decades, particularly in the 1960s when BCIs research was developed [15].

Due to the pioneering work in BCIs, the academia and researchers have started to realize the significance of reading and controlling the brain; consequently, conducting research and proposing new approaches around this domain has been growing rapidly in recent years. Particularly since 2001, efforts have been presented towards materializing the interest of modern science [16]. Furthermore, tech-giants like Facebook or even high-profile entrepreneurs like Elon Musk are intensified new futuristic startups that seek to enhance human capabilities through BCI technology despite its limitations and hindrance. Besides, plenty of neurotechnology-based applications are available now on the market, so they can be used for various non-clinical and research purposes [17], [18].

This paper pursues to present a synopsis of BCI technology,

history, research potential, and review of various approaches. The paper also explores a Mind-Machine Interfaces (MMI) approach called Emotiv Epoc+ Neuroheadset and discusses the experimental findings that aim to identify the emotional parameters of human subjects. The primary contributions of the paper as follows:

- We provide a comprehensive review on three areas of BCI including (i) Brain-Reading Computer, (ii) Brain-Machine Interfaces, and (iii) Mind-Machine Interfaces and different approaches for each area.
- We study potential approaches for each one of the BCI areas and investigate the effectiveness in neurotechnological research.
- We discuss a case study to study six performance metrics of the human brain using Emotiv Epoc+ Neuroheadset and demonstrate the experiments in a lab environment.

The rest of the paper is structured as follows: Section II reviews the relevant literature on BCI and portraits an overview of three research directions. Section III elaborates our contribution by introducing an experimental presentation of Epoc+ Neuroheadsets and evaluations of its performance metrics. The section also draw a discussion on both experimental and survey results. Finally, Section IV concludes the paper.

## II. BRAIN-COMPUTER INTERFACES (BCI)

Brain-Computer Interfaces (BCI) technology is a rapidly developing research field, recognized as an emerging technology, that has attracted researcher's attention from different fields in the last two decades [19]–[21]. BCIs is also known as a class of neurotechnology originally developed for medical assisting applications that may be described as the most sensorial research field due to its connectivity with the human brain [22]–[24]. Crawford et al. [16] addressed BCI as a measurement of the central nervous systems' (CNS) activities, which translate into new CNS outputs without using the brain's normal output channels of peripheral nerves and muscles [4], [25]. Nowadays, researchers are optimistic to develop new augmentative communication and control technology for human patients with severe neuromuscular disorders by adopting BCI technology [26].

The fundamental structure of the modern BCI system is based on four basic components including signal acquisition, signal preprocessing, feature extraction, and classification [28]. Shih et. al. [29] emphasizes the importance of signal-acquisition hardware which must be multi-functional, convenient, portable, safe, adaptable in all environments, and capable to communicate with the brain using electrical signals. According to Guy et al. [30], the first process of BCIs is acquiring brain signals, followed by analyzing them towards decoding the neural activities. Finally, translates the extracting data into commands that are relayed to an output device that carries out the desired action. Instead to use the brain's normal output pathways, these devices measure and use signals produced by the CNS [29]. Fig. 1 Illustrates the low-level architecture of BCI structured introduced by Pfurtscheller et al. [27] that has been widely accepted by many scholars includes [31] and [32]. The first step is to acquire the brain signals by utilizing different methods, for instance, chip-set or electrodes to extract the specific signal features towards classifying the type of intends [33]. Classified features help the operating device to translate into specified commands [27].

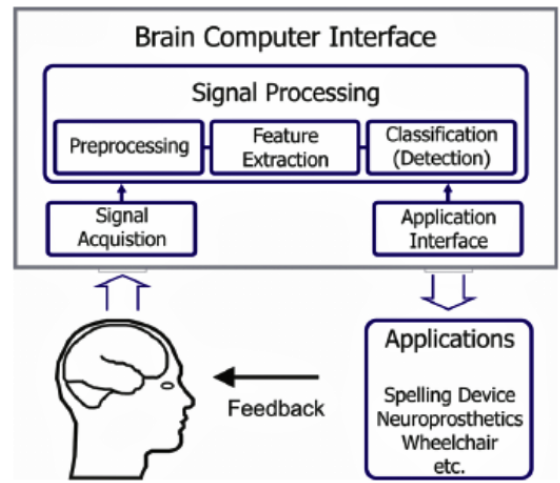

Fig. 1. BCI low-level architecture [27]

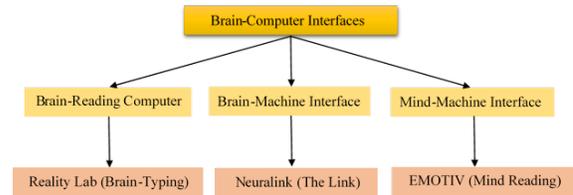

Fig. 2. BCIs based research approaches.

According to a recent update [34], Facebook's Reality Lab (FRL) announced initiation on Brain-Reading Computer (BRC). BRC is a speech decoding approach that allows people to type throughout electrical brain activities without conventional text entry or voice dictation [35]. Besides, Neuralink has unveiled a chip-based Brain-Machine Interface approach, called 'The Link' [36]. The Link aims to introduce a futuristic technology 'Brain-Machine Interface (BMI)' that will enable patients with neurological conditions to control phones or computers with their minds [37]. Other than BRC and BMI, a bio-informatics company, Emotiv, is developing an EEG-based neuroheadset that is capable to use electroencephalography (EEG) and electrodes [38] to read the brain. In this section, we discuss the aforementioned three potential BCIs approaches as shown in Fig. 2. First, we provide an overview of BRC based approach followed by Neuralink's futuristic project, The Link. Finally, we discuss a well know EEG based Mind-Machine model.

## A. Brain-Reading Computer

Facebook's Brain-Computer Interface (BCI) development research team lead by Regina Dugan [39] initiated a hands-free communication (without saying a word) project in 2017 in collaboration with the University of California, San Francisco (UCSF). The research, known as 'Project Steno' mainly focuses on the development of real-time applications based on a brain-reading computer that is capable of decoding speech directly from the human brain onto a screen using machine learning techniques [18]. The primary aim of Project Steno is to let people, particularly patients who have shortness in speaking, type phrases in a computer by thinking about what they wanted to say using a non-invasive, wearable device. This goal pairs with UCSF's research because it intends to improve the lives of people suffering from paralysis and other forms of speech impairment [40].

The model exploits a set of machine learning techniques equipped with refined phonological speech models that can learn and decode the speech using brain activities [41]. It also utilizes recurrent neural networks inspired by state-of-the-art speech recognition and language translation algorithms. [40], [42]. According to a report, [43], the proposed approach, which is still in its infancy, may not decode every thought of a human brain into text, rather it will decode only those speech that would be decided by the person. The decoding process is divided into two segments, 'high-extraction' that is responsible to record the neural activity of brain signals, and 'neural network' that consists of a temporal convolution, an RNN encoder, and a decoder process. The hypothesis has been demonstrated using 30-50 unique sentences. The result shows the promising potential of decoding selective thought of the human brain that is expected to be used for futuristic mind-controlled technology. Researchers from both teams expecting to decode 100 words per minute from 1,000 vocabulary words and an expected decoding error of less than 17 %.

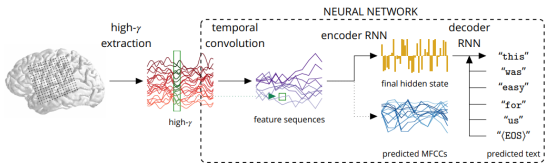

Fig. 3. Process of speech decoding by adopting the neural network approach presented by Moses et al. in 2020 [44].

In a separate study by the Center for Integrative Neuroscience, UCSF, [44] introduced a novel approach that shows robust accuracy of decoding natural speech or typing using electrocorticogram (ECoG) displayed in Fig. 3. Regardless of the demonstration within a limited environment, the researchers show the accuracy of the proposed approach higher than the previous attempts of state-of-the-art phoneme-based classifiers [35]. Approximately 30-50 sentences were used to conduct the experiments of the method, called decoding pipeline. The first step of the method is to record the high-density neural activities of participants. To acquiring data, the researchers utilize Deep Neural Networks (DNNs) techniques where temporal convolutional sequence modeling is used for extracting feature sequences, and the encoding and decoding process used a recurrent neural network (RNN) that predicts the speech and the next word in the sequence respectively.

## B. Brain-Machine Interface

The Brain-Machine Interface (BMI) is intimately related to the effort of developing new electrophysiological methods to record the extracellular electrical activity of large neuronal populations using multi-electrode configurations [45]. For many decades, research in BMI got the attention of researchers globally, especially John Cunningham Lilly [45] who introduced the initial architectural design of BMI in 1950 when he successfully implanted 25–610 electrodes in brain adult rhesus monkeys' brains (See Fig. 4). Modern BMI architecture is designed for both experimental and clinical studies and can translate raw neuronal signals into motor commands [46] as demonstrated by Chapin, J.K. et al. [47]. In this study, researchers proposed an approach that can directly control the cortical neurons using a robotic manipulator.

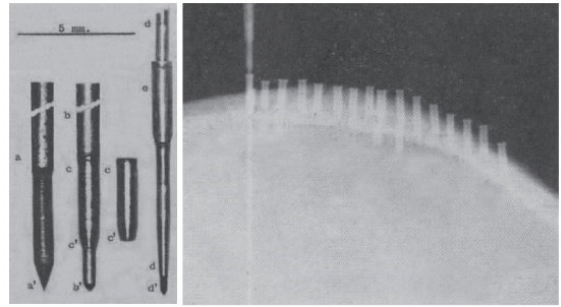

Fig. 4. John Lilly's parts of the electrode implant (left); X-ray of 20 implanted sleeves and one inserted electrode of monkey (right) [45]

A newly established, futuristic company 'Neuralink' announced its visionary project called 'The Link' with the mission of helping people with paralysis to regain independence through the control of computers and mobile devices [48]. Neuralink hopes to use an implantable device to record brain signals and allow people to control computers and other machines with just their thoughts [49]. To describe Neuralink's first steps toward a scalable high-bandwidth BMI system, Elon Musk and Neuralink share [50] the endeavor to develop a visionary product for brain' reading composed of arrays of small and flexible electrode "threads", with as many as 3,072 electrodes per array distributed across 96 threads. In Fig 5 the prototype of the Link is given where letter A refers to the 256 channels data processing unit; B indicates the polymer threads on parylene-c substrate; C is titanium cases that are coated with parylene-c, and D is the power and data connector. The approach of Neuralink, according to its website, [51] is to understand the brain followed by interfacing, and engineering the brain.

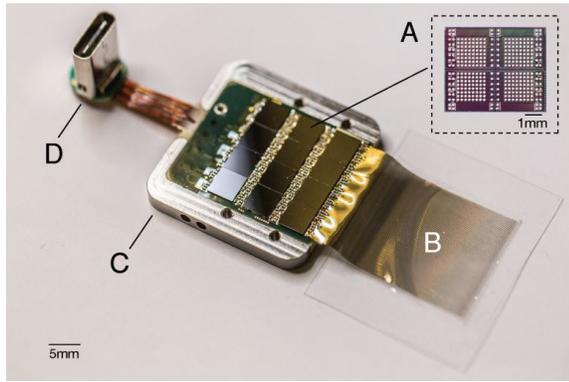

Fig. 5. Architecture of the links' sensor device called Neuralink application-specific integrated circuit (ASIC) [50].

In addition, Neuralink has developed a neurosurgical robot 'robotic electrode inserter' as scratched in Fig. 6 that has the capacity of inserting six threads (192 electrodes) per minute into the brain in an automated mode [50]. In the figure, A indicates the loaded needle pincher cartridge; B shows the brain position sensors while C refers to the light modules consist of multiple independent wavelengths. D, E, F, and G indicate the motor, needle camera, wide-angle view camera, and stereoscopic cameras, respectively. The primary objectives of this robot are to avoid surgery time, vasculature and record from dispersed brain regions. According to Neuralink, the chip must be installed using the automated 'electrode inserter robot' and the electrodes, which will read those impulses, will amplify the signal in the processing unit [52]. In the recent announcement, Elon Musk presents the chips that already implanted in three pig's brain [53]. During the demonstration, Elon Musk shows the reading neural signals of one of the pigs named Gertrude in real-time [54]. The developed approach has exhibited results that achieve a spiking yield of up to 70 % in chronically implanted electrodes [50].

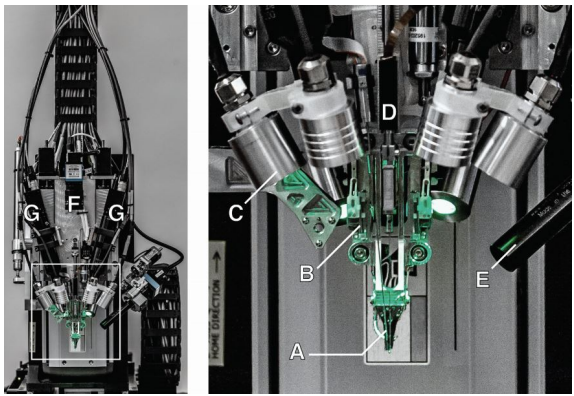

Fig. 6. Robotic electrode Inserter introduced by [50].

### C. Mind-Machine Interface

Professor Eberhard E. Fetz [55] was one of the earliest pioneers on work that connects machines to minds. Since that time, researchers and labs around the world have made impressive progress towards bringing the imagination to reality. Like many others, Emotiv, a bioinformatics company, is focusing to develop varieties of electroencephalography (EEG) based BCIs products with the mission of empowering individuals to understand their brain and accelerate brain research globally since its foundation in 2011 [56]. EEG, according to Williams et al. [14], is one of the oldest neuroscientific techniques and refer to a continuous recording of the electrical activity generated by groups of neurons in the brain. The device calibrates itself by having the subject closing their eyes and remaining in a neutral state of mind without movement. This calibration method provides a relatively clean reading to measure the EEG properly [57]. EEG is capable to read the outer layer of the brain and measure the oscillations of electric impulses in the brain using sensors on the scalp. Emotiv Epoc+ neuroheadset, shown in Fig. 7, is considered as a popular alternative of medical-grade EEG recording devices due to its novelty and affordable commercialization [58], [59]. The Epoc+ headset is considered as an EEG reader that connects with a processing application through Bluetooth technology [60]. Epoc+ headset is capable to detect, read, and acquire data from the user's emotional state, facial expressions using EEG. The device also provides support applications to display the output of the signals as variables [61]. In this sub-section II-C, we review such non-invasive device called Epoc+ Neuroheadset that provides wireless EEG data acquisition and processing.

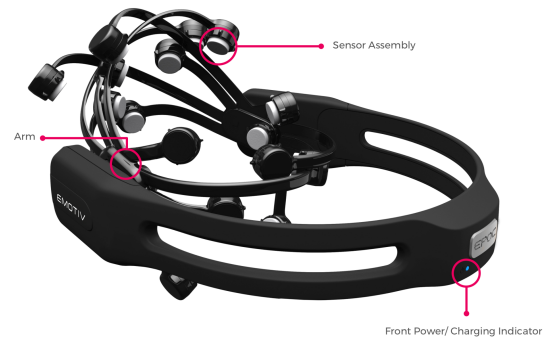

Fig. 7. Physical layout of the Emotiv Epoc+ neuroheadset.

Emotiv neuroheadset (version: Epoc+) consist of 14 Silver-Silver Chloride Electrodes (Ag/AgCl) [62] located at F3, F4, AF3, AF4, F7, and F8 for imaging the lobus frontalis neural activity of the subject's brain; T7, T8, FC5 and FC6 for lobus temporalis while the lobus parietalis is scanned by P8 and P7 electrodes. The other two electrodes O2 and O1 are responsible for the lobus occipitalis in two different arms within 3-axis navigations [38]. Epoc+ reads the data between 128 Hz and 256 Hz and connects with a parent app "Emotiv App" to access the offered features using three other different software applications independently named 'EmotivBCI', 'EmotivPRO', and 'EmotivBrainVIZ' [14]. To evaluate the neuroheadset and its applications, it is necessary to prepare the headset and the applications and adjust the headset with the skull. We found the physical design of the Epoc+ neuroheadset has deficiencies in terms of maintaining the contact quality above 98%, which is required for accessing various features and getting expected readings. Fig. 8 presents the contact quality between the skull and the electrodes during the evaluation that reaches 100 % (left) and 41 % (right). Ensuring contact quality need to be higher in order to access some of features including extraction of 'mental commands' and 'facial expressions'.

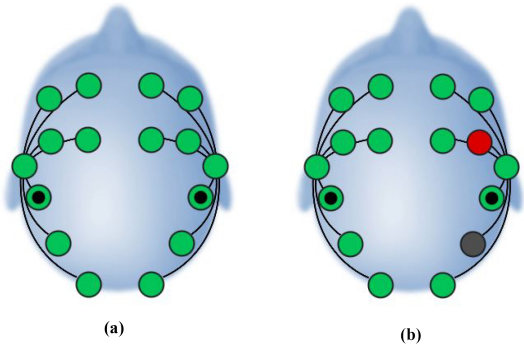

Fig. 8. Overview and percentage of contact quality of Epoc+ Neuroheadset.

According to the researcher Deitz et al. [63], EEG is efficient for measuring rates of decision-making and reading neuro-engagement with dimension. By adopting the concept of EEG, Epoc+ neuroheadset offers scores of features for students and researchers. Besides, Emotiv also offers two exciting features 'Mental Commands' and 'Facial Expression'. Fig. 9 represents the Emotiv training interface of its mental commands and facial expressions. Using the mental commands, the user is allowed to control an object by providing training for different directions, such as 'push', 'lift', and 'rotate right'. Besides, facial expression enables users to repeat the parallel facial expression such as 'smile', 'clench teeth', 'raise brows', etc. However, we demonstrate performance metrics exclusively to investigate and validate the effectiveness of Epoc+ neuroheadset in reading emotions of human subjects. In section III, we discuss the analyzing result of emotional parameters of the participants including (i) engagement, (ii) focus, (iii) excitement, (iv) stress, (v) relaxation, and (vi) interest.

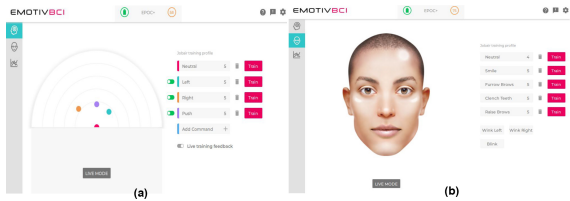

Fig. 9. Emotiv training interface for mental commands (left) and facial expressions (right).

## III. Experiments and Evaluations

### A. Real-Time Datasets

A total of three participants with self-reported healthy conditions have participated in the following experiment. Subject-I and Subject-II is a male of 25 and 62 years old while Subject-III is a female of 48. We first attempted to achieve a connection quality of 100% to investigate the 14 Ag-AgCl electrodes in terms of channel spacing (uV). We suggest the participants instilling their emotions according to the screening video towards detecting their feelings accurately. To prepare the aforementioned participants for the final experiments, we provide necessary training in the lab environment including a briefing about the neuroheadset and an experiment followed by short demo practices align to real-time experiments. To ensure the required connection and reliable values, it is necessary to prepare the headset by hydrating the sensors using multipurpose contact lens saline solution followed by installing the sensors with the headset (each sensor should be inserted into black plastic arms of the headset) that lead to connecting the headset with software applications using Bluetooth. We calibrate the subject using Emotiv App towards getting real-time readings on Emotiv applications. To achieve the expected readings, we 'calibrate' the subject to get a strong signal, we asked the participants to close their eyes and remain in a neutral state followed by keeping their eyes open without blinking. Our implementation commenced by screening different kinds of videos to help participants controlling their emotions by watching given scenarios within the aforementioned categories. The initial result generated by Emotiv app depicted in Fig. 10.

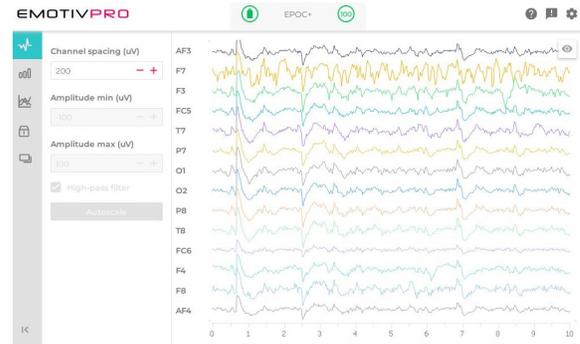

Fig. 10. Illustration of the 200 uV RAW electroencephalography (EEG) diagram for 14 electrodes in 100 % contact quality.

We begin the experiments with subject-I by displaying a movie clip (category-action) on a TV screen for fifteen minutes while we collected the readings for the last eleven minutes and thirty seconds. For the subject-II, we play a math competition video while subject-III (category-attention) was shown a movie clip (category-comedy). We record the readings for an entire segment of experiments and collect a wide range of datasets in CSV format. We extract about 6,086 events (2,066 "subject-I", 2,100 "subject-II", and 1,920 "subject-III") related to the performance metrics from hundred of datasets generated as experimental output by Epoc applications. We split each group of extract datasets into five different categories to monitoring and yielding the emotional states. We also surveillance the flow of each categorized dataset for assessing and validating the output of the performance metrics. Our primary effort is to produce the graphical interfaces using the datasets where the X-axis represents the flow of the dataset and the Y-axis represents the value of the event. We obtain a maximum value of 1.00, 9.016, 131.23, 1.00, 16.672, and 1.049 for Engagement, Excitement, Stress, Relaxation, Interest, and Focus respectively while minimum values are -0.999, 0.00, -2.945, 0.00, 0.00, and 0.00 for the same parameters illustrate (See Figures 11-14 where PM refers to Performance Metrics). During the experiments, Emotiv App allows to monitor the real-time reading of all the parameters on a graphical interface.

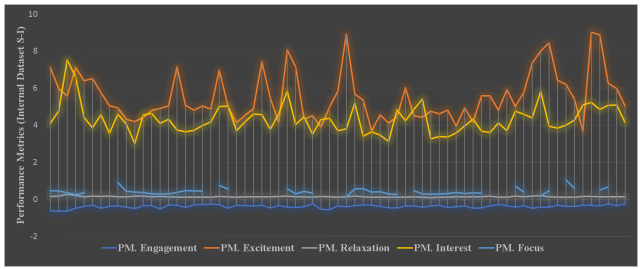

Fig. 11. Overall parameters of performance metrics in the lab environment (100% contact qualities) using internal datasets, subject-I.

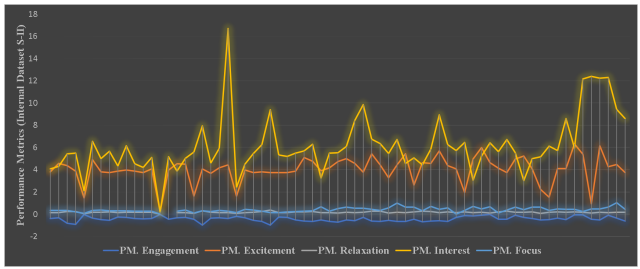

Fig. 12. Overall parameters of performance metrics in the lab environment (100% contact qualities) using internal datasets, subject-II.

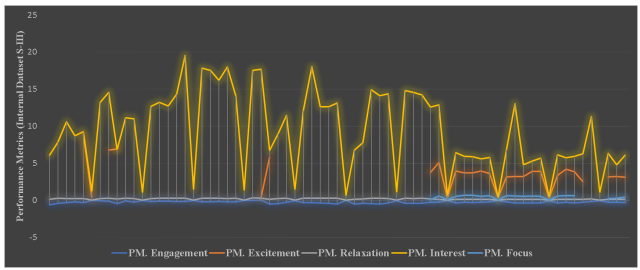

Fig. 13. Overall parameters of performance metrics in the lab environment (100% contact qualities) using internal datasets, subject-III.

During the analysis of hundreds of data, we found missing readings of brain signals for certain times and we assume that the primary cause of absence data of a specific part of the parameters is not having required contact quality. For instance, Figures 11-13 illustrate the paucity of the parameter "Focus", while "Excitement" and "Interest" is seen as the most reading parameters by the electrodes as shown in Fig. 11-12. In subsection III-D, we evaluate the readings to validate the datasets with the preliminary survey in subsection III-C. We attempt to assess possible correlations between the events to determine the viability of EEG's readings.

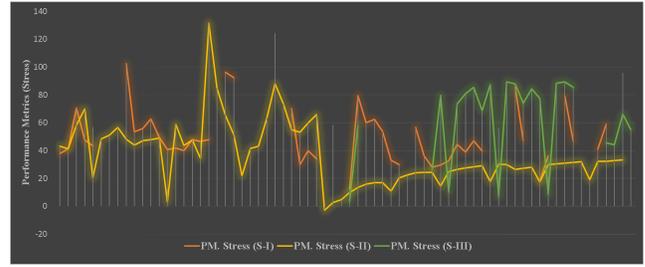

Fig. 14. Parameter (stress) of performance metrics in the lab environment (100% contact qualities) using experimental datasets.

### B. External Datasets

We also collected a vast number of data from Emotiv by adopting quantitative research approach, approximately 202,140 datasets consisting of about 2,170 events of emotional parameters. Our primary goal is to evaluate the events gradually with the real-time events generated from neuroheadset. We identify the minimum values of -1.182, 5.427, 0.886, 3.650, and -1.182 for Engagement, Excitement, Relaxation, Interest, and Focus respectively while the maximum values are -0.130, 16.105, 0.348, 15.549, and 0.844. However, designed values (maximum) for Engagement, Excitement, Relaxation, Interest and Focus are sequentially -0.144, 16.528, 0.462, 17.034, and 1.000 while minimum acceptable values are -1.258, 4.90, -0.010, -5.270, and 0.000 for the aforementioned parameters displayed in Fig. 15-16; X-axis represents the events and Y-axis represents the value of the variables and PM refers to Performance Metrics.

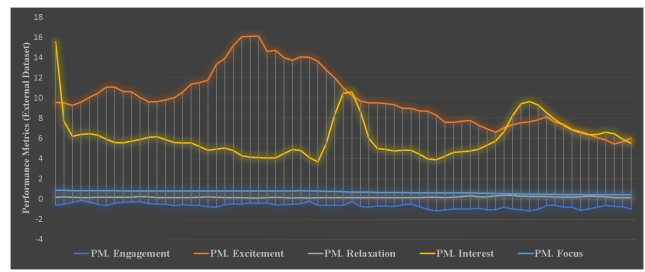

Fig. 15. Overall parameters of performance metrics using external datasets from Emotiv.

### C. Preliminary Survey

We conduct a preliminary survey among all participants of the experimental study to evaluate and validate the readings data. A self-administered questionnaire utilizes to collect the data related to the subject's thoughts where they were asked to answer and complete lists of questionnaires online. We design the survey with a variety of questions including "Rate the specific parameters" on a scale of 1-10 for a particular time-frame when we record the readings.

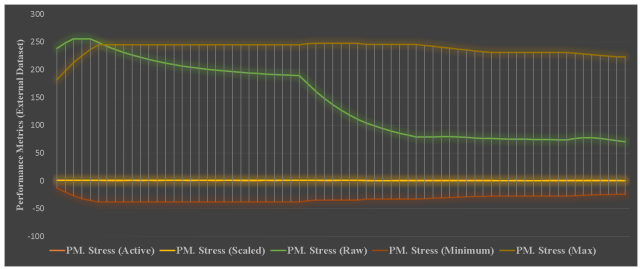

Fig. 16. Parameter (stress) of performance metrics using external datasets from Emotiv.

TABLE I
PRELIMINARY SURVEY: A GENERIC ILLUSTRATION OF THE RATINGS FOR PARAMETERS

| Experiment (First Half) | | | |
|---|---|---|---|
| Parameters | Subject-I | Subject-II | Subject-II |
| Engagement | 0.60 | 0.30 | 0.40 |
| Excitement | 0.30 | 0.70 | 0.80 |
| Stress | 0.20 | 0.60 | 0.30 |
| Relaxation | 0.60 | 0.30 | 0.50 |
| Focus | 0.30 | 0.40 | 0.20 |
| Interest | 0.70 | 0.90 | 0.80 |
| Experiment (Second Half) | | | |
| Parameters | Subject-I | Subject-II | Subject-II |
| Engagement | 0.90 | 0.60 | 0.50 |
| Excitement | 0.70 | 0.90 | 1.00 |
| Stress | 0.70 | 0.60 | 0.10 |
| Relaxation | 0.70 | 0.50 | 0.60 |
| Focus | 0.50 | 0.60 | 0.10 |
| Interest | 0.90 | 0.10 | 1.00 |

For the questionnaires of the survey, we split the time-frame of the experiment into two; (i) first six minutes and (ii) second six minutes of the experiments. We survey the ratings of the subjects for each performance metric during both time-frame as shown in Table I. We also consider asking the subject's basis about the higher rate in two particular performance metrics which are 'Excitement' and 'Interest'.

### D. Evaluation

After conducting training and experiments within a controlled session, we evaluate the findings of performance metrics of Emotiv applications. We aggregate the data to conduct statistical analysis, introduce different but informative diagonals that presents two primary output, (i) correlations between performance metrics (top-right corner) and (ii) histogram of experimental datasets and external datasets (middle) as illustrated in Fig. 17 and Fig. 18 respectively. Both diagrams are unique and we can identify the linear dependency between events (bottom-left corner). The charts also illustrate the correlation between each performance metric.

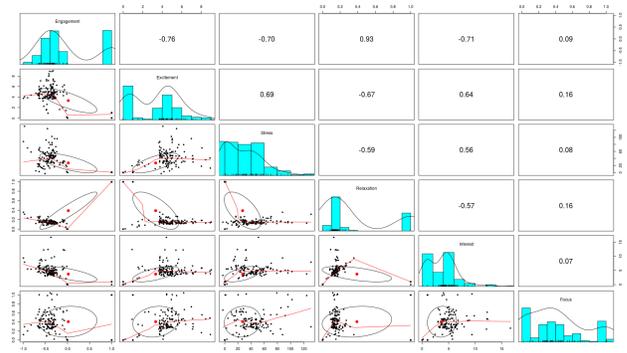

Fig. 17. Probability chart of performance metrics using raw data generated in the lab environment (100% contact qualities).

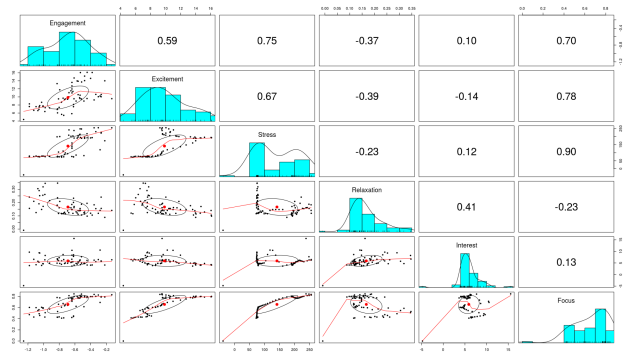

Fig. 18. Probability chart of performance metrics using external datasets collected from Emotiv.

We train and test experimental datasets by adopting two supervised learning algorithms: Naïve Bayes and Linear Regression. We use 70% events for training and 30% for testing purposes from 1,242 events. All of the events splits into six specific parameters in accordance with the Emotiv neuroheadset's readings. For Naïve Bayes, we calculate the probability to obtain the prediction. We first identify the percentage of accurately predict values which is 48% followed by using a confusion matrix that gave an accuracy of 57%. Besides, we identify another statistical accuracy include skewed data measurement named Kappa that gives a primary exactness 32%. Our effort was to increase the exactitude of Kappa that leads the classifier to improve overall accuracy. In our third attempts, we find a considerable realistic accuracy of 69% and Kappa slightly increased to 41%, Table II(a).

In Table II(b), we apply 70% and 30% of overall events for training and testing respectively. Linear regression classifier classified about 800 events and identify 45% and 62% accuracy for two attempts. However, the measured standard error was 15% and the t-Stat was 12%. Besides, demonstration on individual parameters also indicates close accuracy.

To give a comparative representation between the performance metrics of three subjects and their survey report, we conduct a generalized contrastive study. First, we evaluate the associate dataset from experiments, subsequent survey, and extend to external datasets. In a generalizing analysis illustrated in Fig. 19-25 (X-axis and Y-axis represent the event and value of the event respectively), two parameters 'Engagement' and 'Relaxation' of subject-II show gradual steadiness between events. Our participant (Subject-II) affirms possible engagement during the experiments since he was keen to understand the mathematical logic that leads the parameter 'Focus' uncertain reading but higher than other subjects. On the contrary, the other two parameters 'Excitement' and 'Interest' of subject-I show the higher reading and the survey indicates her total interest and excitement in watching movies (action). However, both subject I and II indicates vicissitudes orientation in stress and focus respectively. Subjects agreed that unique experimental techniques, the content of the video clip, and the latest technological equipment used to conduct the experiments were the primary reason for the higher reading of 'Excitement' and 'Interest' parameters. We also evaluated the external datasets, and conclude identical and robust but static flow in each parameters comparing to experimental datasets Fig. 17-22. Although our illustration between three similar datasets, records in different lab environments, participants, and periods, show significant similarity within performance metrics. Moreover, slight differences may be occurred due to disparity between aforesaid datasets of the subject's activities during the experiments.

TABLE II
ANALOGY BETWEEN NAÏVE BAYES AND LINEAR REGRESSION USING
CONFUSION MATRIX (70% TRAINING, 30% TESTING)

| Naïve Bayes (a) | | |
|---|---|---|
| Initial Accuracy | Improved Accuracy | N=1,242 |
| 57% | 69% | Training: 70% |
| 32% (Kappa) | 41% (Kappa) | Testing: 30% |
| Linear Regression (b) | | |
| Initial/Improved Accuracy | Standard Error/t-Stat | N=1,242 |
| 45% | 15% | Training: 70% |
| 62% | 12% | Testing: 30% |

*E. Discussion*

Utilizing our experimental datasets of Epoc+ Neuroheadset, we measure the classification of emotions accuracy 69% and 62% for Naïve Bayes and Linear Regression respectively illustrated in Table II. To optimize the accuracy of Naïve Bayes classification, we adopt the Kappa metric that measures the agreement between classification and truth values; As expected, it increase the classification accuracy by about 9% compare to the first attempt. Within a comparative study between survey and experimental data draw a parallel implication. Besides, we present diagrams for the linear dependency between events of each dataset and complete analogical illustrations. The study also found that preparatory training helps the subjects to complete the experiments intensively. Overall, all dataset (including external) demonstrate the compatibility of the Emotiv approach by reading the EEG signals, converting the data into variables, displaying on the user inter-face, and giving promising accuracy that can be compatible for researchers, neuroscientists, psychologists, rehabilitation specialists around the context of detecting and measuring performance metrics.

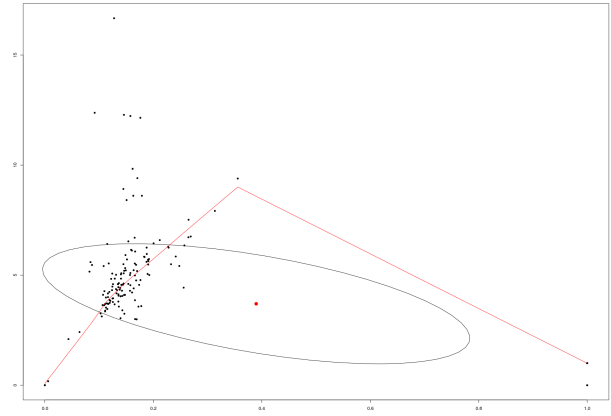

Fig. 19. Graphical illustration between Emotiv and survey datasets for linear regression, X-axis consist of survey data while Y-axis consists of experimental events.

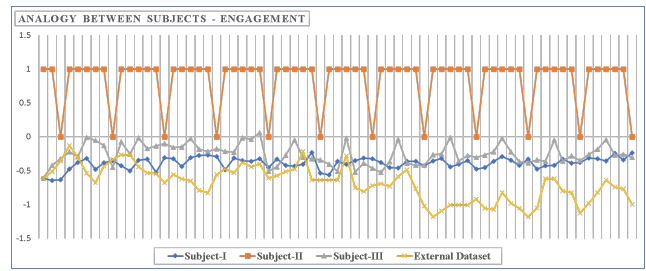

Fig. 20. Analogical illustration based on the parameter Engagement between subjects, 75% experimental and 25% external.

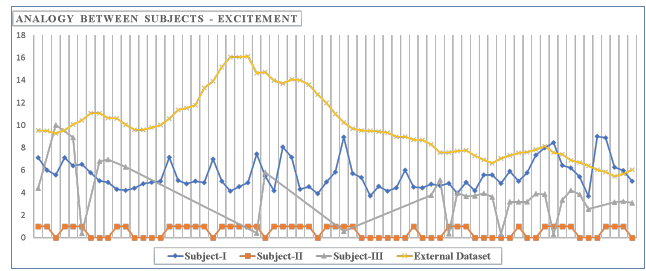

Fig. 21. Analogical illustration based on the parameter Excitement between four subjects, 75% experimental and 25% external.

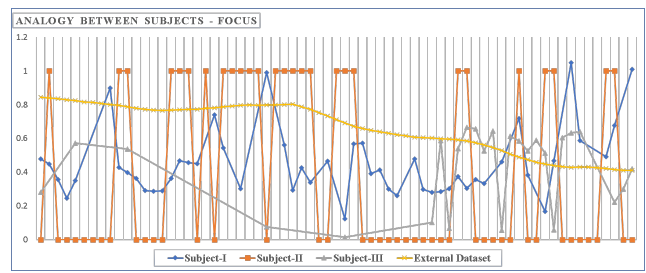

Fig. 22. Analogical illustration based on the parameter Focus between subjects, 75% experimental and 25% external.

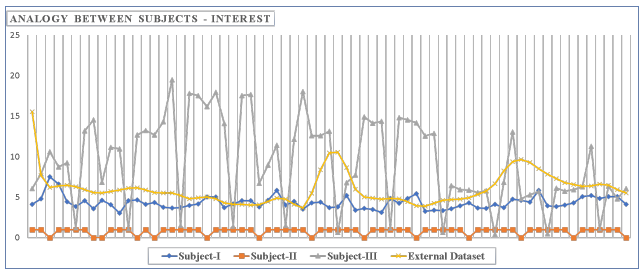

Fig. 23. Analogical illustration based on the parameter Interest between subjects, 75% experimental and 25% external.

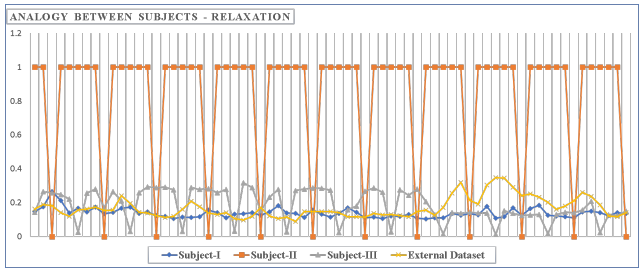

Fig. 24. Analogical illustration based on the parameter Relaxation between subjects, 75% experimental and 25% external.

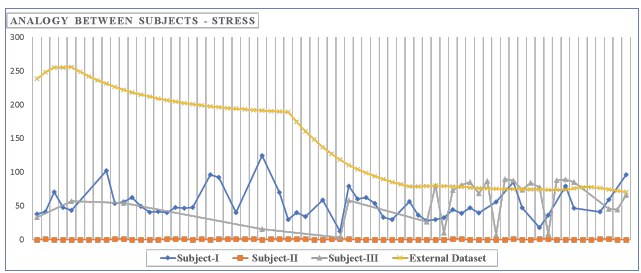

Fig. 25. Analogical illustration based on the parameter Stress between subjects, 75% experimental and 25% external.

## IV. CONCLUSION

Neurotechnology is an untenable research field that has been entrenched to the scholars in academia and BCI research is now a multidisciplinary effort. In this study, we considered investigating three unveiled potential Brain-Computer Interfaces (BCI) research approaches. The paper discussed the FRL and Neuralink's projects in detail and their recently presented demonstration results. Besides, we demonstrated Emotiv Epoc+ Neuroheadset using raw data from three different sources and illustrated its effectiveness by adopting Naïve Bayes and Linear Regression technique towards analyzing and understanding brain activities, electroencephalography (EEG), and Brain-Computer Interfaces (BCI) extensively for future endeavor. We also evaluated the efficiency of the Epoc+ device in terms of reading the performance metrics of human beings with a presentation of findings that indicates acceptance accuracy of the neuroheadset and its applications. So far, experimental findings appear a promising research tool around the field of neurotechnology. Significantly, other approaches revealed their progressional promises in BCIs research towards a technological wonderland. To draw a conclusion, we have not yet considered high-level synthesis, but we aim to continue our effort explicitly towards significant accomplishments around the domain of BCIs; following the vision 2050 when BCI could become a magic wand for developing men control objects with the mind.

## ACKNOWLEDGMENT

This work was supported in part by research computing resources and technical expertise via a partnership between Kennesaw State University's Office of the Vice President for Research and the Office of the CIO and Vice President for Information Technology [64].